\documentclass[10pt,letterpaper,twocolumn]{article} 

\usepackage{ol2}
\usepackage[draft]{hyperref}
\usepackage{amsmath}

\begin{document}

\twocolumn[ 

\title{Localization without recurrence and pseudo-Bloch oscillations in optics}


\author{Stefano Longhi}

\address{Dipartimento di Fisica, Politecnico di Milano and Istituto di Fotonica e Nanotecnologie del Consiglio Nazionale delle Ricerche, Piazza L. da Vinci 32, I-20133 Milano, Italy (stefano.longhi@polimi.it)}

\begin{abstract}
Dynamical localization, i.e. the absence of secular spreading of a quantum or classical wave packet, is usually associated to Hamiltonians with purely point spectrum, i.e. with a normalizable and complete set of eigenstates, which show quasi-periodic dynamics (recurrence). Here we show rather counter-intuitively that dynamical localization can be observed in Hamiltonians with absolutely continuous spectrum, where recurrence effects are forbidden. An optical realization of such an Hamiltonian is proposed based on beam propagation in a self-imaging optical resonator with a phase grating. Localization without recurrence in this system is explained in terms of pseudo-Bloch optical oscillations. 
\end{abstract}



 ] 
  
{\it Introduction.} 
Diffusion, localization and recurrence phenomena play a crucial role in several fields of science \cite{r1}. 
Optics has provided on several occasions an ideal framework to investigate transport and localization effects earlier predicted in quantum physics   \cite{r3,r4}.  In systems described by a Schr\"{o}dinger-like wave equation, general relations between the temporal dynamics and the spectrum of the underlying Hamiltonian have been disclosed (see \cite{r5,r6,r7,r8,r9,r10,r11} and references therein). Roughly speaking, the absolutely continuous spectrum case corresponds to transport and diffusive phenomena, the pure point case
typically corresponds to the absence of transport (localization) and recurrence phenomena, while the singular continuous case corresponds to intermediate transport behavior. Hamiltonians with a purely point spectrum always show almost periodic dynamics owing to the quantum recurrence theorem \cite{r12}, whereas in systems with purely absolutely continuous spectrum recurrence effects are always prevented as a consequence of the Riemann-Lebesgue theorem \cite{r5,r6}. In case of pure point spectrum, in addition to recurrence one {\it generally} observes {\it dynamical localization}, i.e. an initially localized wave packet does not secularly spread, whereas in the absolutely continuous case dynamical delocalization is commonplace. Prominent examples of localization and recurrence associated to a point spectrum are the phenomena of Anderson localization and its ramifications, and of Bloch oscillations, which have been extensively investigated in optics\cite{r13,r14,r15,r16,r17,r18,r18bis,r19,r20,r21,r22,r23,r24,r25,r26,r27,r28,r29}. While the above rule is very common, it is not universal one. For example, dynamical delocalization in systems with pure point spectrum can be found in certain random Hamiltonian models  \cite{r30}.\par
In this Letter we present an example of a quantum  Hamiltonian $\hat{H}$ with  absolutely continuous spectrum that shows dynamical localization, and suggest an optical implementation of this model. The Hamiltonian $\hat{H}$  describes a {\it continuous} extension of the effective Wannier-Stark Hamiltonian in a single band of a crystal \cite{r31}, i.e. it is of the form $\hat{H}=T(\hat{p_x})+Fx$ where the kinetic term $T$ a periodic function of the momentum operator $\hat{p}_x=-i \partial_x$ with period $2 \pi/d$ whereas $x$ is a continuous spatial variable. While in the discrete limit, i.e. by taking $x=nd$ with $n=0, \pm 1, \pm2,...$  the Hamiltonian $\hat{H}$ has a pure  point spectrum (Wannier-Stark ladder), showing localization and quantum recurrence (the famous Bloch oscillations), in the continuous case periodicity is lost according to the Riemann-Lebesgue theorem, however dynamical localization is preserved. We show that such a dynamical behavior, that we may refer to as {\it pseudo-Bloch oscillations}, can be realized in a self-imaging optical ring resonator with an intracavity phase grating. A transition from localization to delocalization by breaking the exact self-imaging condition of the resonator is also discussed.
\par 
{\it Quantum Hamiltonian and pseudo-Bloch oscillations.} Let us consider a quantum system described by the one-dimensional Schr\"{o}dinger equation $ i \partial_t \psi(x,t) = \hat{H} \psi(x,t)$ with a Stark Hamiltonian given by ($\hbar=1$)
\begin{equation}
\hat{H}=T(\hat{p}_x)+Fx,
\end{equation}
where $\hat{p}_x=-i \partial_x$ is the momentum operator, $F$ is the external force and $T(\hat{p}_{x})$ is the effective kinetic energy operator. When the kinetic energy operator is the one of a free particle, i.e. for $T(\hat{p}_x)=\hat{p}_x^2$, the energy spectrum $E$ of $\hat{H}$ is  purely absolutely continuous $-\infty < E < \infty$ with improper eigenfunctions $| \phi_E(x) \rangle$  given by shifted Airy functions. According to the Riemann-Lebesgue theorem, the {\it return probability}, defined as 
\begin{equation}
P_r(t)= | \langle \psi(x,0)| \psi(x,t) \rangle|^2 \equiv \left| \int dx   \psi^*(x,0) \psi(x,t) \right|^2, 
\end{equation}
decays to zero. Moreover, dynamical delocalization occurs. In fact, by changing the reference frame from rest to an accelerated one, the Stark Hamiltonian $\hat{p}_x^2+Fx$ is equivalent to the one of the free particle without the force (see, for example, \cite{r3}). Here we consider the case where $T(q)$ is a period function of $q$ with period $2 \pi/d$, i.e. 
\begin{equation}
T(q)=\sum_{l } T_l \exp(i l d q) 
\end{equation}
with $T_{-l}=T_{l}^*$.  Hamiltonians with a periodic kinetic energy operator are generally introduced in solid-state physics as an effective Hamiltonian to describe single-band electron wave packet dynamics in slowly-varying external fields \cite{r31}. In this case  $T(q)$ is the band dispersion curve, and the discretization $x=nd$ ($n=0, \pm1, \pm2, \pm3,...$) at localized Wannier sites must be accomplished \cite{r31}. As is well-known, with such a discretization $\hat{H}$ has a pure point spectrum with equally-spaced energy levels (Wannier-Stark ladder) and energy separation $\omega_B=Fd$; in real space an initially localized wave packet undergoes a periodic motion, so-called Bloch oscillations, and the return probability $P_r(t)$ is a periodic function with the Bloch oscillation period $T_B=2 \pi / \omega_B=2 \pi/(Fd)$.  Here we consider the case where the spatial variable $x$ is{ \it continuous} rather than discretized. In such a case, it can be shown that $\hat{H}$ has a pure absolutely continuous spectrum $-\infty < E < \infty$ with improper (non-normalizable) eigenfunctions $ | \phi_E(x) \rangle$ given by
\begin{equation}
\phi_E(x) =\sum_n \theta_n \delta(x- nd- E/F)
\end{equation}
where we have set 
\begin{equation}
\theta_n=\frac{d}{2 \pi \sqrt{F}} \int_0^{2 \pi/d} dq \exp \left\{  i q d n+ \frac{i}{F} \int_0^q d \xi  T( \xi) \right\}.
 \end{equation}
 Note that $| \phi_E(x) \rangle$ are not normalizable and satisfy the usual relations of improper eigenfunctions $\langle \phi_{E'} | \phi_E \rangle= \delta(E-E')$. The most general wave packet solution to the Schr\"{o}dinger equation $i \partial_t\psi=\hat{H} \psi$ is given by 
 \begin{equation}
 \psi(x,t)=\int dE \mu(E) \phi_E(x) \exp(-iEt), 
 \end{equation}
 where the spectral measure $\mu(E)$ is determined by the initial wave packet distribution $\psi(x,0)$  according to $\mu(E)= \int dx \phi_E^*(x) \psi(x,0)$. The return probability $P_r(t)$ then reads $P_r(t)= | \int dE |\mu(E)|^2 \exp(iEt)|^2$, which decays to zero as $ t \rightarrow \infty$ according to the Riemann-Lebesgue theorem: hence when the spatial variable $x$ is continuous rather than discrete, the dynamics {\it is not periodic}. Using Eq.(4), Eq.(6) can be cast in the form
 \begin{equation}
 \psi(x,t)=G(x,t) \exp(-iFxt)
 \end{equation}
 where we have set
 \begin{equation}
 G(x,t)=F \sum_{n,l} \theta_l \theta_n^* \psi(x-  ld+ nd,0) \exp(iF d lt)
 \end{equation}
 Note that $G(x,t)$ is a periodic function of $t$ with period $T_B$ and $G(x,0)=\psi(x,0)$. Since $|\psi(x,t)|^2=|G(x,t)|^2$ is periodic, dynamical localization is obtained. However, contrary to ordinary Bloch oscillations the dynamics {\it is not} periodic [this is due to the factor $\exp(-iFxt)$ in Eq.(7)] and the return probability decays to zero. We can refer to such a dynamical behavior, i.e. dynamical localization but absence of recurrence, to as {\it pseudo-Bloch oscillations}. An example of pseudo-Bloch oscillations in an optical resonator will be presented in the following. Finally, let us extend the analysis by considering the Hamiltonian
 \begin{equation}
 \hat{H}=T(\hat{p}_x)+\epsilon \hat{p}_x^2+ Fx,
 \end{equation}
  where $T(q)$ is periodic and defined by Eq.(3) and an additional non-periodic kinetic term of strength $\epsilon$ is included. For $\epsilon \neq0$, it can be shown that pseudo-Bloch oscillations are smeared out and dynamical wave packet delocalization arises. In fact, the solution to the Schr\"{o}dinger equation with Hamiltonian $\hat{H}$ defined by Eq.(9), from time $t_n=nT_B$ to time $t_{n+1}=(n+1)T_B$, is formally given by $\psi(x,t_{n+1})= \exp (-i T_B \hat{H}) \psi(x,t_n)$. By repeated use of the Baker-Campbell-Hausdorff formula and after some cumbersome calculations, it can be shown that $\psi(x,t_{n+1})=\exp(-i T_B \hat{H}_{eff}) \psi(x,t_n)$, where we have set
  \begin{equation}
  \hat{H}_{eff}= \epsilon \hat{p}_x^2+Fx
  \end{equation}
  Such a result, which is exact regardless of the strength of $\epsilon$, shows that the quantum evolution of the Hamiltonian (9), mapped at times $t_n=nT_B$ multiples than the Bloch oscillation cycle, is equivalent to that described by the effective Hamiltonian (10). Since $\hat{H}_{eff}$ is equivalent to the free-particle Hamiltonian after a change of reference frame, it shows dynamical delocalization.
  \par
  
 {\it Optical realization.}  Transverse light dynamics in optical resonators can provide a fertile testbed to emulate continuous quantum mechanical Hamiltonians with engineered kinetic  energy operator \cite{r32,r33,r34}. An optical implementation of the continuous quantum Hamiltonians (1) or (9) with  periodic $T(\hat{p}_x)$ can be obtained by considering the optical resonator shown in Fig.1. It consists of a (near) self-imaging ring cavity of length $L$ in so-called 4$f$ configuration, comprising four equal converging lenses of focal length $f$, one phase grating (PG) and one prismatic phase mask (PM) in the geometrical setting of Fig.1. The total cavity length is $L=8f+l$, with $l \ll f$ and $l=0$ for exact self-imaging. The transmission functions of the PG and of the prismatic PM are given by $t_{1}(x)=\exp[-i \varphi_1(x)]$ and $t_{2}=\exp[-i \varphi_2(x)]$, where $\varphi_1(x+a)=\varphi_1(x)$ is the periodic phase change introduced by the PG with spatial period $a$ and $\varphi_2(x)=i \alpha k x$ is the phase change introduced by the prismatic PM ($k= 2 \pi / \lambda$ is the optical wave number and $\alpha$ is the angle of the prismatic surface). Light propagation inside the optical ring can be readily obtained by application of the generalized Huygens-Fresnel integral (see e.g. \cite{r32,r33,r34}). Assuming one transverse spatial dimension $x$ and indicating by $\psi_m(x)$ the envelope of the intracavity field at the reference plane $\gamma$ in the cavity at the $m-th$ round trip, the following recurrence relation can be derived
 \begin{equation}
 \psi_{m+1}(x)= \hat{\mathcal{D}}_1 \hat{\mathcal{D}}_2 \hat{\mathcal{D}}_3 \psi_m(x)
 \end{equation} 
 \begin{figure}[htb]
\centerline{\includegraphics[width=8.4cm]{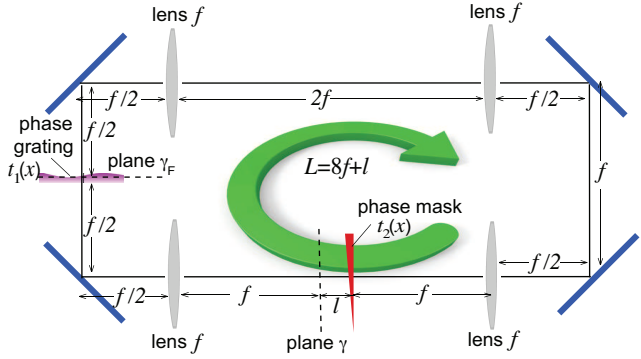}} \caption{ \small
(Color online)  Schematic of the optical ring resonator that realizes pseudo-Bloch optical oscillations. The cavity operates in the exact self-imaging condition for $l=0$. The resonator comprises four lenses (focal length $f$), a phase grating and a prismatic phase mask, with transmittances $t_1(x)=\exp[-i \varphi_1(x)]$ and $t_2(x)=\exp[-i \varphi_2(x)]$ respectively. The total cavity length is $L=8f+l$. Planes $\gamma$ and $\gamma_F$ are Fourier conjugate planes.}
\end{figure}
 where we have set $\hat{\mathcal{D}}_1 \equiv \exp \left(-i \frac{l}{2k} \frac{\partial^2}{\partial x^2} \right)$, $\hat{\mathcal{D}}_2 \equiv \exp[-i \varphi_2(x)]$, and the operator $\hat{\mathcal{D}}_3$ is defined via the relation $\hat{\mathcal{D}_3} f(x) \equiv  \sum_{n} \sigma_n f(x+nd)$, with  $d \equiv  \lambda f/a$ and with coefficients $\sigma_n$ given by the Fourier series $\exp[-i \varphi_1(x)]=\sum_{n} \sigma_n \exp( 2 \pi i n x/a)$.  In writing Eq.(11), we neglected cavity losses and assumed that a single axial mode of the cavity is excited \cite{r32,r34}. Operator $\hat{\mathcal{D}_1}$ on the right hand side of Eq.(11) describes beam diffraction for an effective length $l$ ($l=0$ and $\hat{\mathcal{D}}_1= \mathcal{I}$ for an exact self-imaging cavity), $\hat{\mathcal{D}}_2$ accounts for the additional phase acquired after transmission through the prismatic PM, whereas $\hat{\mathcal{D}}_3$ describes the effect of the PG placed in the Fourier plane $\gamma_F$ (see Fig.1). To clarify the analogy between Eq.(11) and the Hamiltonian (9), let us consider the limits $|\varphi_{1,2}| \ll \pi$ and $l \rightarrow 0$, so that the operators $\hat{\mathcal{D}}_{1,2,3}$ slightly deviate from the identity operator $\mathcal{I}$.  In this case, using standard methods (see e.g. \cite{r32,r33,r34}) after first-order expansion and continuation of the round trip number ($m \rightarrow t$) from Eq.(11) one can derive the following master equation that describes transverse light evolution in the resonator
 \begin{equation}
 i \frac{\partial \psi}{ \partial t}= - \epsilon \frac{\partial^2 \psi}{\partial x^2}+Fx \psi+ \sum_n T_n \psi(x+nd)
  \end{equation}
  \begin{figure}[htb]
\centerline{\includegraphics[width=8.4cm]{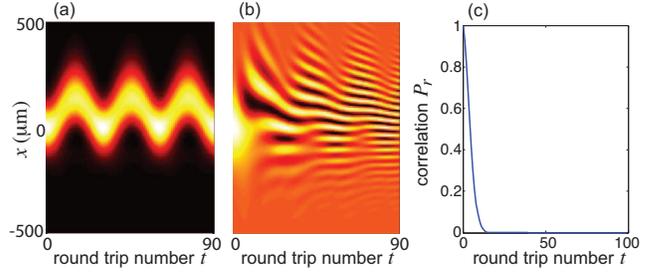}} \caption{ \small
(Color online)  Pseudo-Bloch oscillations in the ring resonator under exact self-imaging condition ($l=0$). (a) and (b) show in pseudo color maps the numerically-computed evolution of the intensity [panel (a)] and real part [panel (b)] of the intracavity field $\psi(x,t)$ at plane $\gamma$ at successive round trips $t$ in the ring. Parameter values are given in the text. The initial condition is the Gaussian field $\psi(x,0) \propto \exp(-x^2/w^2)$ with spot size $w= 150 \; \mu$m. (c) Evolution of the correlation function $P_r(t)$.}
\end{figure}
 \begin{figure}[htb]
\centerline{\includegraphics[width=8.4cm]{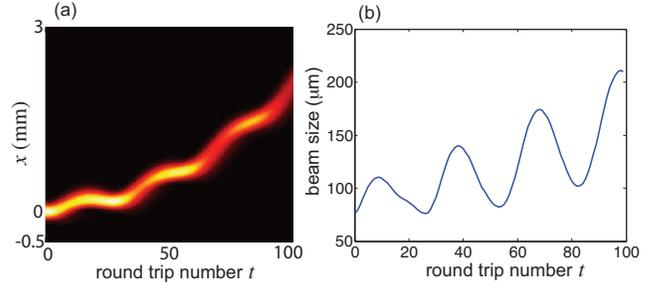}} \caption{ \small
(Color online)  Smearing out of pseudo-Bloch oscillations in the ring cavity under a slight deviation from self-imaging condition ($l=2$ mm). Other parameter values are as in Fig.2. (a) Numerically-computed evolution of the intensity of the intracavity field $\psi(x,t)$ at plane $\gamma$, and (b) of the beam spot size, defined as $[\int dx (x- \langle x \rangle)^2 |\psi(x,t)|^2 / \int dx |\psi(x,t)|^2]^{1/2}$. The secular growth (on average) of the spot size at successive Bloch oscillations cycles is the signature of dynamic delocalization.}
\end{figure}
 where $t$ is the temporal variable in units of the cavity round trip time, $F \equiv k \alpha$, $ \epsilon \equiv -l/(2k)$, and $T_n$ are the Fourier coefficients of the phase $\varphi_1(x)$ \cite{note}. Note that Eq.(12) is precisely the Schr\"{o}dinger equation with Hamiltonian (9). Hence for exact self-imaging condition ($l=0$) the beam evolution over successive round trips in the ring resonator  undergoes pseudo-Bloch oscillations and shows dynamical localization, whereas smearing out of pesudo-Bloch oscillations and delocalization occurs when the resonator length is detuned from exact self-imaging. An example of localization without recurrence arising from pseudo-Bloch oscillations under exact self imaging condition is shown in Fig.2. The figure shows the evolution of an initial Gaussian beam distribution in the ring at successive round trips, as obtained from the propagation map  (11),  for a sinusoidal PG $\varphi_1(x)=\delta \cos(2 \pi x/a)$ and for parameter values $f=3$ cm, $\lambda= 633$ nm, $\delta=0.2$, $a=180 \; \mu$m, $\alpha=0.2 \; {\rm mrad}$ and $L=8f=24 \; {\rm cm}$. Figures 2(a) and (b) show the evolution of the intracavity field intensity $|\psi(x,t)|^2$ and of the real part of the field ${\rm Re}[\psi(x,t)]$, respectively, at plane $\gamma$; in (c) the evolution of the correlation function $P_r(t)=| \int dx \psi^*(x,0) \psi(x,t)|^2$ (the analogue of the return probability) is  also depicted. Note that, while the intensity distribution undergoes a periodic dynamics with period $T_B= 2 \pi /(Fd)= a/( \alpha f) \simeq 30$ round trips, the dynamics is not periodic and the field does not reproduces itself, as one can see from Figs.2(b) and (c).\\
 Smearing out of pseudo-Bloch oscillations and dynamical delocalization are observed when the ring deviates from the self-imaging condition. As an example, Fig.3 shows the beam dynamics for the same conditions of Fig.2, except for an effective diffraction length $l=2$ mm in the ring. In this case one clearly observes that the oscillatory dynamics of Bloch oscillations is superimposed to a transverse acceleration of the optical beam [Fig.3(a)]  and a spreading of the beam size on average [Fig.3(b)]. Such a behavior arises from the fact that the beam dynamics mapped at successive Bloch oscillation cycles is described by the effective Hamiltonian (10). Given the form of $\hat{H}_{eff}$, effective diffraction and transverse index gradient are responsible for spreading and transverse beam acceleration.\par
 To experimentally observe pseudo-Bloch oscillations, one can consider the passive ring resonator of Fig.1 driven by an holding beam coupled through one of the four mirrors of the cavity (see also Ref.\cite{r34}). Assuming that the frequency of the holding beam is in resonance with one of the cavity axial modes, the map (11) is replaced by the following one
 \begin{equation}
 \psi_{m+1}(x)= \hat{\mathcal{D}}_1 \hat{\mathcal{D}}_2 \hat{\mathcal{D}}_3 \psi_m(x)+\sqrt{T} E_m(x)-\frac{T}{2} \psi_m(x),
 \end{equation}
 where $T \ll1$ is the transmittance of the coupling mirror and $E_m(x)$ is the spatial profile of the injected beam at the $m$-th round trip and at plane $\gamma$ ($E_m(x)$ is a slowly-varying function of $m$). Similarly, taking into account cavity losses and external beam injection the master equation (12) is modified as follows
 \begin{equation}
 i \frac{\partial \psi}{ \partial t}= - \epsilon \frac{\partial^2 \psi}{\partial x^2}+Fx \psi+ \sum_n T_n \psi(x+nd)-i\frac{T}{2} \psi+i\sqrt{T} E(x,t).
  \end{equation}  
Pesudo-Bloch oscillations can be observed when a Gaussian holding beam is switched off and the transient decay dynamics of the passive ring resonator (i.e. the transverse beam distribution over several round-trips during the transient decay) are monitored outside the cavity by a fast CCD camera, as discussed in details in Ref.\cite{r34}. As an example, Figs.4(a) and (b) show the numerically-computed evolution of the normalized intensity distributions of the field $\psi(x,t)$ at plane $\gamma$ after switching off the Gaussian holding beam $E(x,t)=f(t) \exp(-x^2/w^2)$ with a decaying amplitude  $f(t)=(1/2)\{1-{\rm tanh}[(t-t_0) / \Delta t] \}$ [dashed curve in Fig.4(c)]; parameter values used in the simulations are $\lambda=633$ nm, $f=3$ cm, $\alpha=0.1 \; {\rm mrad}$, a sinusoidal PG $\varphi_1(x)=\delta \cos(2 \pi x/a)$ with $\delta=0.3$ and $a=180\; \mu$m, output coupling  $T=1 \%$, spot size of the holding Gaussian beam $w=180 \; \mu$m, and switch off time $\Delta t=8$ (in units of round-trip time). In Fig.4(a) exact self-imaging of the cavity ($L=8f=24 \; {\rm cm}$) is assumed, whereas in (b) a slight deviation from self-imaging by $l=1$ mm is  assumed. The solid curve in Fig.4(c) shows the evolution of optical power available outside the resonator during the switch-off dynamics, normalized to its continuous-wave value before switching off the holding beam.  Note that, before the holding beam is switched off, a stationary intracavity field distribution is established, which deviates from the Gaussian one because of the effects of PG, transverse index gradient and [in (b)] diffraction. As the holding beam is switched off at round trip time $t_0=200$, the dynamical behaviors of beam intensity in Figs.4(a) and (b) become similar to those shown in Figs.2(a) and 3(a), respectively. Note also that, since the Bloch oscillation period is smaller than the photon lifetime in the cavity, a few Bloch oscillation cycles can be detected during the decay dynamics.\\ 
 \\

  \begin{figure}[htb]
\centerline{\includegraphics[width=8.4cm]{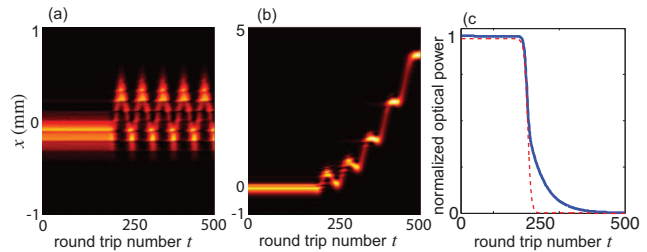}} \caption{ \small
(Color online)  Pseudo-Bloch oscillations (color maps of normalized intensity distribution at plane $\gamma$) in the ring resonator with a Gaussian holding beam for (a) exact self-imaging, and (b) for a detuned cavity ($l=1$ mm). Parameter values are given in the text. Panel (c) shows the behavior of normalized optical power (solid line) and amplitude $f(t)$ of the Gaussian holding beam (dashed line) during the switch-off time.}
\end{figure}
 \par
{\it Conclusions.} 
 In conclusion, the spreading properties of quantum or classical wave packets described by a Schr\"{o}dinger-like equation are known to be closely related to the spectral properties of the Hamiltonian. An absolutely continuous spectrum is generally associated to dynamical delocalization and absence of recurrence, whereas a pure point spectrum generally corresponds to localization and recurrence. However, such rules are not universal. Here we have shown rather counter-intuitively that dynamical localization can be observed in Stark Hamiltonians with purely continuous spectrum, and introduced the novel phenomenon of pseudo-Bloch oscillations. An optical realization of the quantum Hamiltonian, based on transverse beam dynamics in a ring resonator, has been suggested. 
Our analysis indicates that light dynamics in optical resonators can provide a testbed for the observation of other unusual localization and diffusion processes. For example, by proper engineering the power law of effective diffraction in the resonator \cite{r33}, regimes like quasi-absence of diffusion predicted in certain quantum models \cite{r9} could be observed in optics.

\par


\footnotesize {\bf References with full titles}\\
\\
1. P. Sheng, {\it Introduction to Wave Scattering, Localization, and Mesoscopic Phenomena} (Springer, 2006).\\
2. S. Longhi, {\it "Quantum-optical analogies using photonic structures"}, Laser \& Photon. Rev.
{\bf 3}, 243 (2009).\\
3. M. Segev, Y. Silberberg, and D.N. Christodoulides, {\it "Anderson localization of light"}, Nature Photon. {\bf 7}, 197 (2013).\\
4. C.R. de Oliveira, {\it Intermediate Spectral Theory and Quantum Dynamics} (Birkh\"{a}user, Basel, 2008).\\
5. G. Teschl, {\it Mathematical Methods in Quantum Mechanics} (AMS, 2009).\\
6. S. Fishman, D. R. Grempel, and R. E. Prange, "{\it Chaos, Quantum Recurrences, and Anderson Localization}", Phys. Rev. Lett. {\bf 49}, 509 (1982).\\
7. R. Ketzmerick, G. Petschel, and T. Geisel, "{\it Slow Decay of Temporal Correlations in Quantum Systems with Cantor Spectra}", Phys. Rev. Lett. {\bf 69}, 695 (1992).\\
8. F. Cannata and L. Ferrari, "{\it Anomalous diffusion in quantum systems with absolutely continuous spectra}", J. Phys. I France {\bf 1}, 1029 (1991).\\
9. J. Bellissard and H. Schulz-Baldes, "{\it Subdiffusive Quantum Transport for 3D Hamiltonians with Absolutely Continuous Spectra}", J. Stat. Phys. {\bf 99}, 587 (2000).\\
10. Y. Otobe and I. Sasaki, "{\it  Measure theoretical approach to recurrent properties for quantum dynamics}", J. Phys. A {\bf 44}, 465209 (2011).\\
11. P. Bocchieri and A. Loinger, "{\it Quantum Recurrence Theorem}", Phys. Rev. {\bf 107}, 337 (1957).\\
12. D. S. Wiersma, P. Bartolini, A. Lagendijk, and R. Righini, "{\it  Localization of light in a disordered medium}", Nature {\bf 390}, 671 (1997).\\
13. B. Fischer,  A. Rosen, and S. Fishman, {\it "Localization in frequency for periodically kicked light propagation in a dispersive single-mode fiber}, Opt. Lett. {\bf 24}, 1463 (1999).\\
14. B. Fischer, B. Vodonos, S. Atkins, and A. Bekker, {\it "Experimental demonstration of localization in the frequency domain of mode-locked lasers with dispersion"},
 Opt. Lett. {\bf 27}, 1061 (2002).\\
15. T. Pertsch, U. Peschel, J. Kobelke, K. Schuster, H. Bartelt, S. Nolte, A. T\"{u}nnermann, and F. Lederer, {\it "Nonlinearity and Disorder in Fiber Arrays"}, Phys. Rev. Lett. {\bf 93}, 053901 (2004).\\
16. T. Schwartz, G. Bartal, S. Fishman, and M. Segev, "{\it Transport and Anderson localization in disordered two-dimensional photonic lattices}", Nature {\bf 446}, 52 (2007).\\
17. Y. Lahini, A. Avidan, F. Pozzi, M. Sorel, R. Morandotti, D. N. Christodoulides, and Y. Silberberg, "{\it Anderson Localization and Nonlinearity in One-Dimensional Disordered Photonic Lattices}", Phys. Rev. Lett. {\bf 100}, 013906 (2008).\\
18. Y. Lahini, R. Pugatch, F. Pozzi, M. Sorel, R. Morandotti, N. Davidson, and Y. Silberberg, "{\it Observation of a localization transition in quasiperiodic photonic lattices"}, Phys. Rev. Lett. {\bf 103}, 013901 (2009).\\
19. S. Longhi, "{\it Metal-insulator transition in the spectrum of a frequency-modulation mode-locked laser}", Phys. Rev. A {\bf 77}, 015807 (2008).\\
20. C. Conti and A. Fratalocchi, {\it "Dynamic light diffusion, three-dimensional Anderson localization and lasing in inverted opals"},  Nature Phys. {\bf 4}, 794 (2008).\\
21. V. Folli and C. Conti, "{\it Self-induced transparency and the Anderson localization of light"}, Opt. Lett. {\bf 36}, 2830 (2011).\\
22. V. Folli and C. Conti, "{\it Anderson localization in nonlocal nonlinear media"}, Opt. Lett. {\bf 37}, 332 (2012).\\
23. S. Karbasi, C.R. Mirr, P. G. Yarandi, R.J. Frazier, K.W. Koch, and A. Mafi, {\it "Observation of transverse Anderson localization in an optical fiber"}, Opt. Lett. {\bf 37}, 2304 (2012).\\
24. S. St\"{u}tzer, Y. V. Kartashov, V. A. Vysloukh, A. T\"{u}nnermann, S. Nolte, M. Lewenstein, L. Torner, and A. Szameit, "{\it Anderson cross-localization}", Opt. Lett. {\bf 37}, 1715 (2012).\\ 
25. O.V. Borovkova, V.E. Lobanov, Y.V. Kartashov, V.A. Vysloukh, and L. Torner, {\it "Dynamic versus Anderson wave-packet localization"}, 
Phys. Rev. A {\bf 91}, 063825  (2015).\\
26. U. Peschel, T. Pertsch, and F. Lederer, {\it "Optical Bloch oscillations in waveguide arrays"}, Opt. Lett. {\bf 23}, 1701 (1998).\\
27. R. Morandotti, U. Peschel, J. S. Aitchison, H. S. Eisenberg, and Y. Silberberg, {\it "Experimental Observation of Linear and Nonlinear Optical Bloch Oscillations"},
Phys. Rev. Lett. {\bf 83}, 4756 (1999).\\  
28. T. Pertsch, P. Dannberg, W. Elflein, A. Br\"{a}uer, and F. Lederer, {\it "Optical Bloch Oscillations in Temperature Tuned Waveguide Arrays"}, 
Phys. Rev. Lett. {\bf 83}, 4752 (1999).\\
29.  N. Chiodo, G. Della Valle, R. Osellame, S. Longhi, G. Cerullo, R. Ramponi, P. Laporta, and U. Morgner, {\it "Imaging of Bloch oscillations in erbium-doped curved waveguide arrays"}, Opt. Lett. {\bf 31}, 1651 (2006).\\
30. C.R. de Oliveira and R.A. Prado, {\it "Dynamical Delocalization for the one-dimensional Bernoulli Discrete Dirac Operator"}, J. Phys. A {\bf 38}, L115 (2005).\\
31.  J.M. Ziman, {\it Principles of the Theory of Solids} (Cambridge University Press, 1965, first ed.), pp. 147?153.\\
32. S. Longhi, {\it "Phase transitions in Wick-rotated PT-symmetric optics"}, Ann. Phys. {\bf 360}, 160 (2015).\\
33. S. Longhi, {\it "Fractional Schr\"{o}dinger equation in optics"}, Opt. Lett. {\bf 40}, 1117 (2015).\\
34. S. Longhi, {\it "Synthetic gauge fields for light beams in optical resonators"}, Opt. Lett. {\bf 40}, 2941 (2015).\\
35. Note that, in the limit $|\varphi_1(x)| \ll \pi$, one can write $\exp[-i \varphi_1(x)] \simeq 1-i \varphi_1(x)$ and hence $\sigma_n=\delta_{n,0}-iT_n$, where $T_n$ are the Fourier coefficients of the phase $\varphi_1(x)$, i.e. $\varphi_1(x)=\sum_n T_n \exp(2 \pi i n x/a)$.


\noindent 


 




   
\begin{thebibliography}{99}



\bibitem{r1}
P. Sheng, {\it Introduction to Wave Scattering, Localization, and Mesoscopic Phenomena} (Springer, 2006).
\bibitem{r3}
 S. Longhi,  Laser \& Photon. Rev. {\bf 3}, 243 (2009).
 \bibitem{r4}
M. Segev, Y. Silberberg, and D.N. Christodoulides, Nature Photon. {\bf 7}, 197 (2013).
\bibitem{r5}
C.R. de Oliveira, {\it Intermediate Spectral Theory and Quantum Dynamics} (Birkh\"{a}user, Basel, 2008).
\bibitem{r6}
G. Teschl, {\it Mathematical Methods in Quantum Mechanics} (AMS, 2009).
\bibitem{r7}
S. Fishman, D. R. Grempel, and R. E. Prange,Phys. Rev. Lett. {\bf 49}, 509 (1982).
\bibitem{r8}
 R. Ketzmerick, G. Petschel, and T. Geisel, Phys. Rev. Lett. {\bf 69}, 695 (1992).
 \bibitem{r9}
F. Cannata and L. Ferrari, J. Phys. I France {\bf 1}, 1029 (1991).
\bibitem{r10}
J. Bellissard and H. Schulz-Baldes, J. Stat. Phys. {\bf 99}, 587 (2000).
\bibitem{r11}
Y. Otobe and I. Sasaki, J. Phys. A {\bf 44}, 465209 (2011).
\bibitem{r12}
P. Bocchieri and A. Loinger, Phys. Rev. {\bf 107}, 337 (1957).
 \bibitem{r13}
D. S. Wiersma, P. Bartolini, A. Lagendijk, and R. Righini, Nature {\bf 390}, 671 (1997).
\bibitem{r14}
 B. Fischer,  A. Rosen, and S. Fishman,  Opt. Lett. {\bf 24}, 1463 (1999).
 \bibitem{r15}
 B. Fischer, B. Vodonos, S. Atkins, and A. Bekker, Opt. Lett. {\bf 27}, 1061 (2002).
 \bibitem{r16}
T. Pertsch, U. Peschel, J. Kobelke, K. Schuster, H. Bartelt, S. Nolte, A. T\"{u}nnermann, and F. Lederer, Phys. Rev. Lett. {\bf 93}, 053901 (2004).
\bibitem{r17}
T. Schwartz, G. Bartal, S. Fishman, and M. Segev, Nature {\bf 446}, 52 (2007).
\bibitem{r18}
Y. Lahini, A. Avidan, F. Pozzi, M. Sorel, R. Morandotti, D. N. Christodoulides, and Y. Silberberg, Phys. Rev. Lett. {\bf 100}, 013906 (2008).
\bibitem{r18bis}
Y. Lahini, R. Pugatch, F. Pozzi, M. Sorel, R. Morandotti, N. Davidson, and Y. Silberberg, Phys. Rev. Lett. {\bf 103}, 013901 (2009).
\bibitem{r19}
 S. Longhi, Phys. Rev. A {\bf 77}, 015807 (2008).
 \bibitem{r20}
C. Conti and A. Fratalocchi, Nature Phys. {\bf 4}, 794 (2008).
\bibitem{r21}
V. Folli and C. Conti, Opt. Lett. {\bf 36}, 2830 (2011).
\bibitem{r22}
V. Folli and C. Conti, Opt. Lett. {\bf 37}, 332 (2012).
\bibitem{r23}
S. Karbasi, C.R. Mirr, P. G. Yarandi, R.J. Frazier, K.W. Koch, and A. Mafi, Opt. Lett. {\bf 37}, 2304 (2012).
\bibitem{r24}
S. St\"{u}tzer, Y. V. Kartashov, V. A. Vysloukh, A. T\"{u}nnermann, S. Nolte, M. Lewenstein, L. Torner, and A. Szameit, Opt. Lett. {\bf 37}, 1715 (2012).
\bibitem{r25}
 O.V. Borovkova, V.E. Lobanov, Y.V. Kartashov, V.A. Vysloukh, and L. Torner,  Phys. Rev. A {\bf 91}, 063825  (2015).
 \bibitem{r26}
 U. Peschel, T. Pertsch, and F. Lederer, Opt. Lett. {\bf 23}, 1701 (1998).
 \bibitem{r27}
R. Morandotti, U. Peschel, J. S. Aitchison, H. S. Eisenberg, and Y. Silberberg,  Phys. Rev. Lett. {\bf 83}, 4756 (1999).
\bibitem{r28}
T. Pertsch, P. Dannberg, W. Elflein, A. Br\"{a}uer, and F. Lederer,  Phys. Rev. Lett. {\bf 83}, 4752 (1999).
\bibitem{r29}
 N. Chiodo, G. Della Valle, R. Osellame, S. Longhi, G. Cerullo, R. Ramponi, P. Laporta, and U. Morgner, Opt. Lett. {\bf 31}, 1651 (2006).
 \bibitem{r30}
 C.R. de Oliveira and R.A. Prado, J. Phys. A {\bf 38}, L115 (2005).
 \bibitem{r31}
 J.M. Ziman, {\it Principles of the Theory of Solids} (Cambridge University Press, 1965, first ed.), pp. 147-153.
\bibitem{r32}
S. Longhi,  Ann. Phys. {\bf 360}, 160 (2015).
\bibitem{r33}
S. Longhi, Opt. Lett. {\bf 40}, 1117 (2015).
\bibitem{r34}
S. Longhi, Opt. Lett. {\bf 40}, 2941 (2015).
\bibitem{note}
Note that, in the limit $|\varphi_1(x)| \ll \pi$, one can write $\exp[-i \varphi_1(x)] \simeq 1-i \varphi_1(x)$ and hence $\sigma_n=\delta_{n,0}-iT_n$, where $T_n$ are the Fourier coefficients of the phase $\varphi_1(x)$, i.e. $\varphi_1(x)=\sum_n T_n \exp(2 \pi i n x/a)$.


\end{thebibliography}
\end{document}